\documentclass[aps,prev,preprintnumbers,floatfix,nofootinbib,12pt]{revtex4-1}
\usepackage{graphicx}
\usepackage{bm}
\usepackage{times}
\usepackage{slashed}
\usepackage{color}
\newcommand{\be}{\begin{equation}}
\newcommand{\ee}{\end{equation}}
\newcommand{\beq}{\begin{eqnarray}}
\newcommand{\eeq}{\end{eqnarray}}
\newcommand{\ba}{\begin{array}}
\newcommand{\ea}{\end{array}}

\begin{document}

\title{A Brief Review on WIMPs in 331 Electroweak Gauge Models}

\author{P. S. Rodrigues da Silva}\email{psilva@fisica.ufpb.br}

\affiliation{Departamento de F\'{\i}sica, Universidade Federal da Para\'{\i}ba, Caixa Postal 5008, 58051-970, Jo\~ao Pessoa, PB, Brasil.
}

\date{\today}
\vspace{1cm}

\begin{abstract}
In this work we review the cold dark matter problem in the context of a class of models which are a simple extension of the electroweak standard model, where the gauge symmetry is dictated by the group structure, $SU(3)_c\otimes SU(3)_L\otimes U(1)_X$. This model, in different versions, has been able to address many interesting facts not explained by the standard model. It would be just desirable that the dark matter issue could be contemplated as well, since the problem of the missing matter has become one of the greatest indications of incompleteness of currently established theories describing the known interactions and the cosmological evolution and content of the observed Universe. We do that by pointing out some of the successful steps in this direction and remarking some sources of difficulties concerning their phenomenological and theoretical implementation.
\end{abstract}
\maketitle



\section{Introduction} \setcounter{equation}{0}
\label{sec1}
The last decades have seen an unprecedented accumulation of astrophysical and cosmological evidence towards the missing matter paradigm, culminating in 
recent data from the Planck satellite~\cite{planck} that have improved our knowledge of matter content of the Universe, revealing that a non-baryonic dark matter amounts to roughly $27\%$ in contrast to $23\%$ of WMAP data~\cite{wmap}, both in the context of the standard cosmological model with a cosmological constant, $\Lambda$. 
That lead us to fiercely conclude that there exists some new component (or components) in the energy content of the Universe which must correspond to some exotic particle in the framework of a particle physics model. If this is the right track, and that is the view we adopt in this review, such a particle should present certain features to explain its stability, abundance, non-relativistic behavior at decoupling from thermal bath and evasiveness while offering the possibility of being detected soon (not mandatory but desirable). One of the most attractive possibilities is the so called weakly interacting massive particle (WIMP),  a particle with mass in the few GeV scale (even a TeV) that interacts most like a neutrino does, thus gathering the right properties to fulfill these requirements to explain this unknown matter, usually called cold dark matter (CDM).

Simple extensions to the electroweak standard model (EWSM) can deal with this unsolved puzzle, like adding a singlet fermionic field, plus a symmetry to guarantee its stability~\cite{singlet}. Although simplicity may seem the appropriate path to attack any new riddle we may face, we look at such models more like a guide, hoping that something more fundamental and ambitious may be hiding under this cloak. This leads us to look for more intricate models available  that might shed light on other questions not embraced by the EWSM, but can also deal with the CDM problem. They are many (see for instance Ref.~\cite{DMmodels,Murayama,dmoutros,Dobrescu,little} and references therein), and each one poses a new content or mechanism that opens up new phenomenological possibilities that can be tested in current, as the Large Hadron Collider (LHC), and future colliders. Here we focus on a specific class of models, which represents only a small increase in the electroweak gauge group compared to the EWSM, namely, the $SU(3)_c\otimes SU(3)_L\otimes U(1)_X$ gauge model, or 331 for short.

Some of the interesting features of these models are, among others, that anomaly cancellation and QCD asymptotic freedom demand that only three family of fermions should be present in the theory, explaining the family problem~\cite{PleitezPisano};  electric charge quantization is automatic in these models~\cite{qcharge}; the tiny neutrino mass can be naturally explained~\cite{lightnu} in a version of the model where the right handed neutrino is in the same multiplet along with its partners that form the usual doublet under EWSM~\cite{331valle,331RH,singer}, or even considering effective operators in the minimal model where the third leptonic component is the right handed charged lepton~\cite{massnu}; the strong CP problem through Peccei-Quinn mechanism can be implemented, yielding a nonthermal candidate for the CDM problem, the axion~\cite{axion}; the Higgs physics probed by LHC can be easily accommodated and also allow for some room to new Physics phenomena~\cite{331Higgs}; the discrepancy between theory and experiment on $(g-2)_\mu$ can be accounted for in some versions of the model~\cite{qcharge,g-2mu}. Besides, it is possible for some versions of 331 model to present a natural CDM candidate as a WIMP~\cite{jcap,331LHNUS}, and it is this specific topic in the context of 331 models that we want to discuss in this review.

This work is organized as follows: In section~\ref{sec2} we present the models which are the focus of this review; Next, in section~\ref{sec3}, we discuss the emergence of a WIMP in two specific versions of 331 models, basically arguing about the role of symmetries in protecting the WIMP from decaying and commenting about the results and additional complementary studies concerning these WIMPs in the context of these models; we then present some final remarks in section~\ref{sec4}.

\section{The 331 models with neutral fermions}
\label{sec2}
There are basically two kinds of 331 models which differ in the electric charge of the third component of the $SU(3)_L$ fundamental representation. One possibility is that this component possesses one unit of electric charge~\cite{PleitezPisano} and the other is when it is electrically neutral~\cite{331valle,331RH,singer}. We are not considering the first possibility in this review as it does not seem to allow for any reasonable symmetry that the claimed CDM candidate is stable or metastable~\cite{dm331sem}. The other possibility, with a neutral fermion as the third component of the lepton triplet, splits into two alternatives: the first is to put the right handed partner of the EWSM neutrino, $(\nu^c)_L$, as the third component of the leptonic triplet; the second is to introduce a new left handed neutral fermion, $N_L$, instead. The 331 model with a right handed partner for the EWSM active neutrino we call 331RH$\nu$ from now on, while the 331 model with a new left handed neutral fermion we call  331LHN .
We present these two possibilities below and highlight their differences~\footnote{There is also the Simple 331 model~\cite{simple}, but we are not considering it here either once it was shown to be ruled out by electrowek precision data~\cite{331dead}.}.

\subsection{331RH$\nu$ and 331LHN Content}
\label{ssec1}

In the 331RH$\nu$ model the leptons are arranged in triplet and in singlet representations of $SU(3)_L$ as follows,
\begin{eqnarray}
f_{aL} = \left (
\begin{array}{c}
\nu_{aL} \\
e_{aL} \\
(\nu_{aR})^C
\end{array}
\right )\sim(1\,,\,3\,,\,-1/3)\,,\,\,\,e_{aR}\,\sim(1,1,-1),
 \end{eqnarray}
where the index $a$ labels the three known families, $a = 1,\,2,\,3$, and the transformation rule under the gauge group is indicated in parentheses. Concerning the quark sector, we choose the first two families in an anti-triplet representation of $SU(3)_L$, while the third one transforms as a triplet with the following content,
\begin{eqnarray}
&&Q_{iL} = \left (
\begin{array}{c}
d_{i} \\
-u_{i} \\
d^{\prime}_{i}
\end{array}
\right )_L\sim(3\,,\,\bar{3}\,,\,0)\,,u_{iR}\,\sim(3,1,2/3),\,\,\,\nonumber \\
&&\,\,d_{iR}\,\sim(3,1,-1/3)\,,\,\,\,\, d^{\prime}_{iR}\,\sim(3,1,-1/3),\nonumber \\
&&Q_{3L} = \left (
\begin{array}{c}
u_{3} \\
d_{3} \\
u^{\prime}_{3}
\end{array}
\right )_L\sim(3\,,\,3\,,\,1/3),u_{3R}\,\sim(3,1,2/3),\nonumber \\
&&\,\,d_{3R}\,\sim(3,1,-1/3)\,,\,u^{\prime}_{3R}\,\sim(3,1,2/3)\,,
\label{quarks} 
\end{eqnarray}
where the index $i=1,2$ refers to the first two generations. The primed fields
are new heavy quarks, actually, leptoquarks since they also carry lepton number~\footnote{This property is a trademark in 331 models due to the fact that these quarks interact with the new gauge bosons that carry lepton number.}.

In order to engender the electroweak symmetry breaking, generating the gauge boson and                                                               fermion masses, we need a minimum of three scalar triplets~\footnote{A version with only two scalar triplets exists~\cite{eco331}, but like those models in Ref.~\cite{dm331sem}, it has not been presented a symmetry to stabilize the CDM candidate. Besides, its lifetime is smaller than necessary to be compatible with a good CDM candidate~\cite{wimplife}, unless a huge fine tuning is adopted to make it reasonable. }, whose field distribution is,
\begin{eqnarray}
 \chi = \left (
\begin{array}{c}
\chi^0 \\
\chi^{-} \\
\chi^{\prime 0}
\end{array}
\right ),\,
\eta = \left (
\begin{array}{c}
\eta^0 \\
\eta^- \\
\eta^{\prime 0}
\end{array}
\right ),\,\rho = \left (
\begin{array}{c}
\rho^+ \\
\rho^0 \\
\rho^{\prime +}
\end{array}
\right ) 
, \label{scalarcont} 
\end{eqnarray}
with $\eta$ and $\chi$ both transforming as $(1\,,\,3\,,\,-1/3)$
and $\rho$ transforming as $(1\,,\,3\,,\,2/3)$ under the 331 symmetry. These scalars are supposed to acquire VEV in the following fashion, in order to ignite the spontaneous symmetry breaking without promoting the lepton number spontaneous violation,
\begin{eqnarray}
 \eta^0 , \rho^0 , \chi^{\prime 0} \rightarrow  \frac{1}{\sqrt{2}} (v_{\eta ,\rho ,\chi^{\prime}} 
+R_{ \eta ,\rho ,\chi^{\prime}} +iI_{\eta ,\rho ,\chi^{\prime}}). 
\label{vacua} 
\end{eqnarray} 

The 331LHN model presents almost exactly the same content except for a small change. The leptonic content is instead,
\begin{eqnarray}
f_{aL} = \left (
\begin{array}{c}
\nu_{aL} \\
e_{aL} \\
N_{aL}
\end{array}
\right )\sim(1\,,\,3\,,\,-1/3)\,,\,\,\,
\nonumber \\
e_{aR}\,\sim(1,1,-1)\,,\,\,\,N_{aR}\,\sim(1,1,0),
 \end{eqnarray}
where we have just changed the right handed neutrino, $(\nu_{aR})^C$,  by a left handed neutral fermion, $N_{aL}$, besides adding its singlet right handed counterpart, $N_{aR}$, to the spectrum.
Nothing changes in the quark content. It may seem an insignificant change but we will see that there are implications for the CDM phenomenology.

A simplified Yukawa lagrangian and scalar potential is obtained in these models by imposing the following discrete symmetry:
\beq
&&\left( \chi\,,\,\rho\,,e_{aR}\,,\, u_{aR}\,,\,u^{\prime}_{3R}\,,\,d^{\prime}_{iR}\,,\, Q_{3L} \right) \rightarrow 
-\left( \chi\,,\,\rho\,,e_{aR}\,,\, u_{aR}\,,\,u^{\prime}_{3R}\,,\,d^{\prime}_{iR}\,,\, Q_{3L}\right)\,.
	\label{z2sym}
\eeq
The most general scalar potential can then be written as,
\begin{eqnarray} V(\eta,\rho,\chi)&=&\mu_\chi^2 \chi^2 +\mu_\eta^2\eta^2
+\mu_\rho^2\rho^2+\lambda_1\chi^4 +\lambda_2\eta^4
+\lambda_3\rho^4+ \nonumber \\
&&\lambda_4(\chi^{\dagger}\chi)(\eta^{\dagger}\eta)
+\lambda_5(\chi^{\dagger}\chi)(\rho^{\dagger}\rho)+\lambda_6
(\eta^{\dagger}\eta)(\rho^{\dagger}\rho)+ \nonumber \\
&&\lambda_7(\chi^{\dagger}\eta)(\eta^{\dagger}\chi)
+\lambda_8(\chi^{\dagger}\rho)(\rho^{\dagger}\chi)+\lambda_9
(\eta^{\dagger}\rho)(\rho^{\dagger}\eta) \nonumber \\
&&-\frac{f}{\sqrt{2}}\epsilon^{ijk}\eta_i \rho_j \chi_k +\mbox{H.c}\,,
\label{potential}
\end{eqnarray}
which is the same for both models, while the Yukawa interactions for the 331RH$\nu$ model writes,
\begin{eqnarray}
&-&{\cal L}^Y =f_{ij} \bar Q_{iL}\chi^* d^{\prime}_{jR} +f_{33} \bar Q_{3L}\chi u^{\prime}_{3R} + g_{ia}\bar Q_{iL}\eta^* d_{aR} \nonumber \\
&&+h_{3a} \bar Q_{3L}\eta u_{aR} +g_{3a}\bar Q_{3L}\rho d_{aR}+h_{ia}\bar Q_{iL}\rho^* u_{aR}+ G_{aa}\bar f_{aL} \rho e_{aR}+\mbox{H.c}\,. 
\label{yukawa}
\end{eqnarray}
The 331LHN model, on the other hand, has one additional term to be included in the Yukawa lagrangian~\footnote{A Majorana mass term for the right handed neutral fermion could be included as well, since it is sterile,  but that is not relevant for the CDM analysis.},
\be
-{\cal L}^Y \supset +g^{\prime}_{ab}\bar{f}_{aL}\chi N_{bR}+\mbox{h.c.}\,.
\label{yukawa2}
\ee
This set of Yukawa interactions, after spontaneous symmetry breaking, allows for mass terms to all fermions, except the active neutrinos which, it may be assumed, obtain their masses through effective dimension-five operators according to Ref.~\cite{lightnu}.

Concerning the gauge sector, both models recover the usual SM gauge bosons, $W^{\pm}\,,\,Z^0\,,\, \gamma$,  and contain five additional vector bosons,  $V^{\pm}$, $U^0$, $U^{0 \dagger}$ and $Z^{\prime}$~\cite{331RH}, with masses around  1~TeV. The kinetic terms involving these gauge bosons imply new neutral and charged currents (not shown here) that, together with the Yukawa couplings in Eqs.~(\ref{yukawa}) and (\ref{yukawa2}), reveal an interesting feature of some of the new fields, they carry two units of lepton number\footnote{Depending on the nature of the third component of the leptonic triplet, this assignment can change to one unit when this component carries no lepton number~\cite{dongsoa,3311,wimpyDR}.},
\begin{eqnarray} {\mbox {\bf L}}(V^+\,,\, U^{\dagger0}\,,\, u^{\prime}_{3} \,,\, \eta^{\prime
0}\,,\,\rho^{\prime +})=-2 \,,\,\,\,\,\,{\mbox {\bf L}}(V^- \,,\,U^0 \,,\,d^\prime_{i}
\,,\, \chi^ 0\,,\, \chi^-)=+2. \label{leptonnumber} \end{eqnarray}
This assignment is such that the lagrangian is lepton number conserving. 

\subsection{Mass spectrum}
\label{spectrum}

With the lagrangian defined we can determine the mass spectrum of these models. As already discussed, all fermions acquire mass from the Yukawa terms in Eqs.~(\ref{yukawa}) and (\ref{yukawa2}) or higher dimension effective operators (the standard neutrinos). Let us focus here on the new neutral fermions of 331LHN model. In the absence of a Majorana mass term for the $N_{L/R}$, Eq.(\ref{yukawa2}) gives only a Dirac mass for these fermions which is,
\be
m_{D}= \frac{g^{\prime}_{ab}}{\sqrt{2}}v_{\chi^{\prime}}\,.
\label{mN}
\ee
At times, it may be appropriate to add a Majorana mass term for these fermions though, $\frac{M}{2} \bar{N_{bR}^c} N_{bR}$. This is the case of Ref.~\cite{wimpyDR}, where the neutral fermion undergoes a seesaw mechanism and the heavier component plays the role of a mother particle that decays into a CDM particle during the radiation era to mimic relativistic species that may be showing up in several cosmological data (see for example Ref.~\cite{planck} and references therein).
In this case we would have two mass eigenstates,
\begin{equation}
N^{\prime}_L = N_L + \frac{m_D}{M} N_R^c \ \,\,\,\,\,\,\,\,\,\mbox{and}\,\,\,\,\,\,\,\,\, N^{\prime}_R = N_R + \frac{m_D}{M} N_L^c,
\end{equation}
with respective eigenvalues,
\begin{equation}
M_{N^{\prime}_L} = \frac{m_D^2}{M}\ \,\,\,\,\,\,\,\,\,\mbox{and}\,\,\,\,\,\,\,\,\, M_{N^{\prime}_R} = M.
\label{massesSEESAW}
\end{equation}
By supposing that the Majorana mass emerges at very high energy scale, $ M >> m_D$, $N_L$ becomes the light component and $N_R$ the heavy one.

Concerning the new quarks, according to Eq.~(\ref{yukawa}) their masses are proportional to the heavy VEV, $v_{\chi^\prime}$, and we assume they are heavier enough (few TeV) so as to assure that one of the new neutral fields can be a stable CDM candidate in the model.

The extra gauge bosons have masses,
\begin{eqnarray}
m^2_{V}    &=& m^2_{U^0} = \frac{1}{4}g^2(v_{\chi^\prime}^2+v^2)\,,
\nonumber \\
m^2_{Z^\prime} &=& \frac{g^{2}}{4(3-4s_W^2)}[4c^{2}_{W}v_{\chi^\prime}^2 +\frac{v^{2}}{c^{2}_{W}}+\frac{v^{2}(1-2s^{2}_{W})^2}{c^{2}_{W}}]\,,
\label{massvec}
\end{eqnarray}
where $g$ is the standard electroweak gauge coupling, $s_W$ (and $c_W$) refers to the Sine (and Cosine) of the electroweak mixing angle and $v=\frac{v_\eta}{2}=\frac{v_\rho}{2}$ is the EWSM breaking scale.

Given the approximations made over the parameters to obtain analytic expressions for the massive CP-even scalar eigenstates, described in Refs.~\cite{jcap,331LHNUS}, we have,
\begin{eqnarray}
H = \frac{(R_\eta +R_\rho)}{\sqrt{2}},\,S_1 =R_{\chi^{\prime}} ,\, S_2 = \frac{(R_\eta - R_\rho)}{\sqrt{2}}\,,
\end{eqnarray}
with the respective eigenvalues,
\begin{eqnarray}
M_{H}  &=&  \sqrt{ 3\lambda_{2}}v\,, \nonumber \\
M_{S_{1}} & = & \sqrt{ \frac{v^{2}}{4}+2\lambda_{1}v_{\chi^\prime}^{2}}\,, \nonumber \\
M_{S_{2}} & = &\sqrt{ \frac{1}{2}\left( v_{\chi^\prime}^{2}+2v^{2}(2\lambda_{2}-\lambda_{6})\right)}\,. 
\label{massashiggs}
\end{eqnarray}
Here, $H$ is identified with the recently detected Higgs boson~\cite{ATLASCMS}, which is the lightest CP-even eigenstate.
There is one massive CP-odd scalar field whose mass eigenstate is,
\be
P_1 \approx \frac{v}{v_{\chi^\prime}}I_{\chi^\prime} + \frac{(I_\eta + I_\rho)}{\sqrt{2}}\,,
\label{cp-odd}
\ee
with the corresponding mass eigenvalue,
\be
M_{P_{1}} =  \sqrt{ \frac{1}{2}(v_{\chi^\prime}^{2} +\frac{v^{2}}{2})}\,.
\label{masscp-odd}
\ee

Besides the Goldstone bosons absorbed by the massive gauge fields, and not shown here, we also have one last neutral complex scalar whose mass eigenstate is,
\be
\phi \approx \left(\frac{v}{v_\chi^\prime}\chi^{*0}+\eta^{\prime 0}\right)\,,
\label{phi}
\ee
and its respective eigenvalue,
\be
m_{\phi} =	\sqrt{ \frac{(\lambda_{7} + \frac{1}{2} )}{2}[v^{2}+v_{\chi^\prime}^{2}] }\,.
\label{phimass}
\ee

Finally, the charged scalars are given by the following mass eigenstates,
\begin{eqnarray}
h^-_1 & = & \frac{1}{(1+\frac{v^2}{v_{\chi^\prime}^2})}(\frac{v}{v_{\chi^\prime}}\chi^{-} + \rho^{\prime-})\,, \nonumber \\
h^-_2 & = & \frac{1}{\sqrt{2}}(\eta^{-} + \rho^{-} )\,,
\label{chargedeigenvectors}
\end{eqnarray}
whose masses are, respectively,,
\begin{eqnarray}
M^{2}_{h^{-}_{1}} & = & \frac{\lambda_{8}+\frac{1}{2} }{2}(v^{2}+v_{\chi^\prime}^{2})\,, \nonumber \\
M^{2}_{h^{-}_{2}} & = & \frac{v_{\chi^\prime}^{2}}{2}+\lambda_{9}v^{2}\,.
\label{massash1h2}
\end{eqnarray}

This completes the whole mass spectrum that is necessary to select the CDM particle in the 331 models here reviewed. Next we analyze the viable candidates considering peculiarities of each version of 331 models with neutral fermions in the leptonic triplets.

\section{CDM in 331 models}
\label{sec3}
The particle physics explanation to the CDM problem demands a particle (or many) that possibly does not interact with most of the SM particles, if any. If this interaction happens to be similar in strength to the weak interaction, and the candidate is stable (or meta-stable) with mass in the range of few GeV to some TeV, it is called WIMP (weakly interacting massive particle). Besides fulfilling these requirements, such WIMP has to produce the correct abundance as inferred by the cosmic microwave background radiation (CMB) power spectrum as measured by WMAP and Planck satellites~\cite{wmap,planck} . The combined data of these experiments result in the CDM abundance, within $68\%$~C.L.,
\be
0.1172 < \Omega_{DM}h^2 < 0.1226\,. 
\label{limitsDM}
\ee
Finally, current direct detection experiments have reported no convincing WIMP signature~\cite{dama,cdms2,cogent}, putting bounds on the maximum WIMP-nucleon scattering cross section~\cite{wimpexperiments,cdms2,xenon100,LUX}. Thus, besides building a model that yields a WIMP that does not overpopulate the Universe, considering the upper bound in Eq.~(\ref{limitsDM}), it is imperative that the parameters producing the appropriate thermal averaged  WIMP annihilation cross section also comply with the upper bounds from detection experiments. Next we discuss the two models, 331RH$\nu$ and 331LHN under the light of the WIMP paradigm.

\subsection{WIMP in 331RH$\nu$}

In the 331RH$\nu$ model of Ref.~\cite{jcap}, the particle that accomplishes these criteria is the excitation of the lightest bilepton field, which for an appropriate range of parameter space, turns out to be the scalar, $\phi$. 
This particular choice was made based on the fact that this scalar can only couple to another bilepton, and if it is the lightest of its kind, lepton number conservation, together with the assumed $Z_2$ symmetry in Eq.~(\ref{z2sym}), would guarantee its stability. Notice that according to Eq.~(\ref{yukawa}), there is no WIMP-lepton coupling since lepton number is assumed to be conserved by the tree level lagrangian and the vacuum.
The authors have shown that there is room in the parameter space for a WIMP that gives the right abundance and is in agreement with direct detection searches. However, the claimed $Z_2$ symmetry that precludes some terms in the scalar potential  is spontaneously broken, since the neutral components of $\chi$ and $\rho$ fields acquire VEV, which could jeopardize this scheme through unwanted operators leading to an unstable $\phi$. It happens that, even after the spontaneous breaking of $Z_2$, there is an apparent symmetry that operates on the bilepton scalars, since $\phi$ only couples to other bileptons, keeping the WIMP stable at tree level~\footnote{  
Effective dimension five operators were supposed to generate neutrinos masses in the model~\cite{jcap,lightnu}, suppressed by high grand unification scale though, $\Lambda_{GUT}\simeq 10^{14}$~GeV.  One of them gives a Dirac mass term for the neutrinos,
\[
\frac{y^{\prime\prime}}{\Lambda_{GUT}}(\bar{f_{aL}^C}\chi^*)(\eta^\dagger f_{bL})+\mbox{h.c.}\,,
\label{efop1}
\]
and would imply the fast WIMP decay into two neutrinos with a lifetime roughly about few seconds. This operator explicitly breaks the $Z_2$ symmetry in the model though and does not represent any danger to WIMP stability.}.
Actually,  when lepton number is a global symmetry it may be broken by gravitational effects and dangerous operators suppressed by Planck scale, $\Lambda_{Pl}\approx 10^{19}$~GeV, may arise to destroy the WIMP stability~\cite{singletportal} . Nevertheless, we cannot envisage any effective operator that would lead to a WIMP lifetime smaller than the allowed limit, $\tau \gtrsim 10^{26-30}$~s~\cite{wimplife}. By naively considering the most dangerous possible effective operator~\footnote{The authors in Ref.~\cite{tuly} wrote down the terms in the scalar potential that violate lepton number at tree level, which are not present in the potential Eq.~(\ref{potential}) due to (spontaneously broken) $Z_2$ symmetry in Eq.~(\ref{z2sym}), but would reappear here when combined with another $Z_2$ violating term.},
\beq
\frac{1}{\Lambda_{Pl}^2}\epsilon_{ijk}\overline{(f_{aL})_i }(f_{bL}^C)_j \rho^*_k \chi^\dagger \eta +{\mbox h.c.}\,,
\label{deop}
\eeq
we would conclude that the WIMP lifetime would be of the order of $\tau\simeq 10^{43}$s, higher enough to consider this WIMP as stable.

\begin{figure}[h]
\centering
\includegraphics[width=0.7\columnwidth]{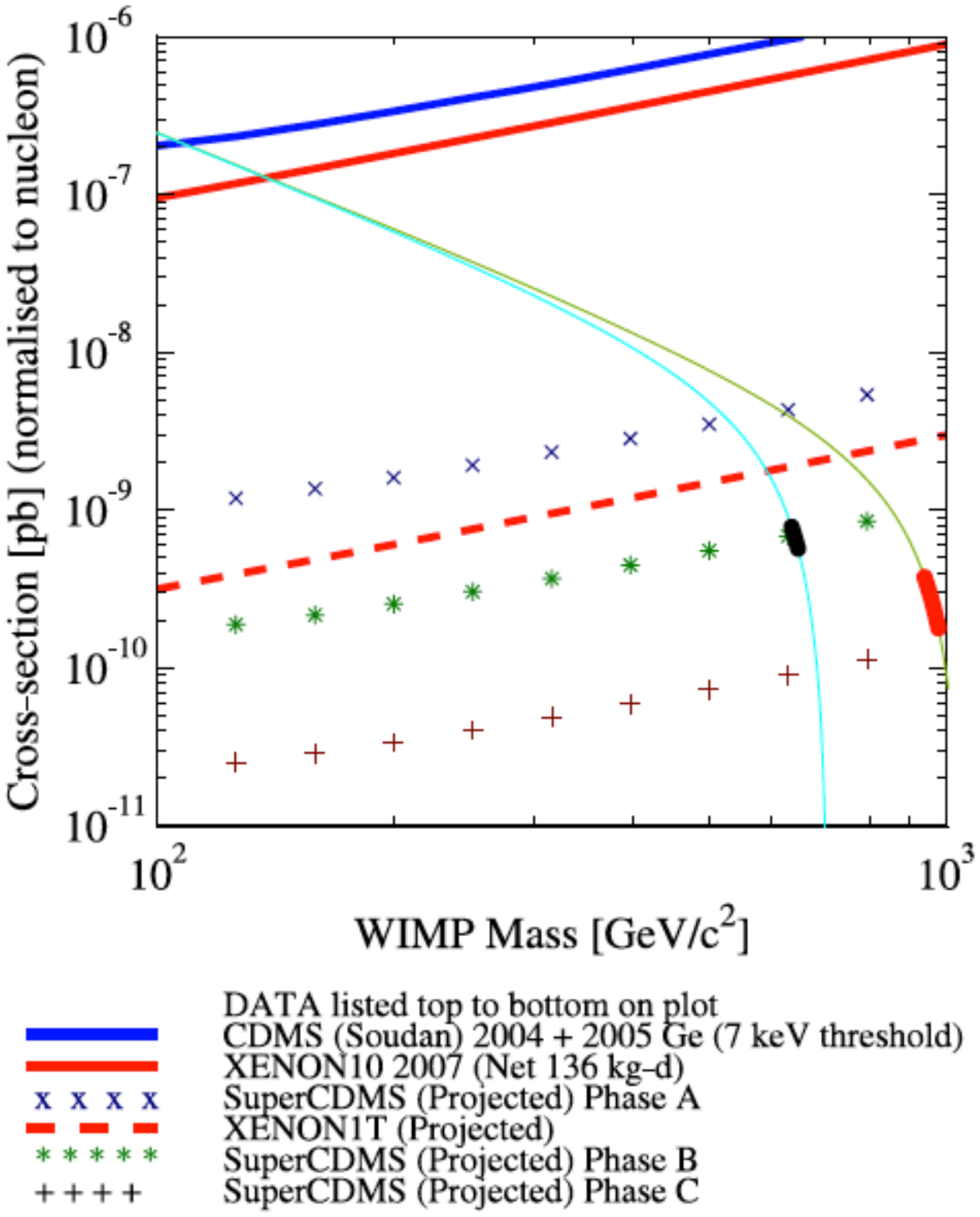}
\caption{Spin-independent WIMP-nucleon elastic scattering cross section sensitivity (current and projected) from CDMS and XENON Collaborations. The two thin lines crossing the data represent the SI WIMP-nucleon cross section of the 3-3-1$RH\nu$ model as a function of the WIMP mass.The upper curve is for $v_{\chi^\prime}=3$~TeV and the lower one is for $v_{\chi^\prime}=2$~TeV, both taken for $M_{q^\prime}=1.5$~TeV. The thicker regions on these lines are those in agreement with the bounds on $\Omega_{CDM}$ imposed by WMAP 3-year run. Figure extracted from Ref.~\cite{jcap}.}
\label{fig:1}
\end{figure}

The model was tested under the light of WMAP-3 year run results~\cite{WMAP3}, concerning the relic CDM abundance, and direct detection experiments, CDMS (2004+2005) and XENON10~\cite{wimpexperiments},  besides some projected experiments~\cite{xenon100,SuperCDMS,xenon1t}. The analysis was rather crude, since several of the parameters of the model were fixed instead of using a scan~\footnote{This drawback was to be fixed only later in a study of the 331LHN model~\cite{FJMZL}.}. Also, the relic abundance was computed by using the approximate analytic  Boltzmann equation~\cite{DMmodels,KolbTurner}. Nevertheless, the results have shown that there exists a restrict range of parameters guaranteeing that the scalar $\phi$ is the lightest bilepton and also a good WIMP candidate, furnishing the correct abundance and safe from direct detection, as presented in Fig.~\ref{fig:1}. 
Basically, a WIMP of 600~GeV to 1~TeV was probed in the 331RH$\nu$ model and the concordance with relic abundance, as measured by WMAP, demanded a $SU(3)_L\otimes U(1)_N$ breaking scale, $v_{\chi^\prime} \gtrsim 1.3$~TeV.
Those results are still out of the sensitivity range of current direct detection experiments, as is the case of XENON100~\cite{xenon100}, but may be tested soon with augmented detector sensitivity~\cite{xenon1t,LZ}. We can infer that a WIMP mass higher than 1~TeV may emerge from an update once bounds on the $Z^\prime$ mass imply a higher value $v_{\chi^\prime} \gtrsim 5.5$~TeV~\cite{yara}.

\subsection{WIMP in 331LHN}

Another variation with a second neutral fermion in the 331 leptonic triplet, that allows for a WIMP in the spectrum was considered in Ref.~\cite{331LHNUS}, the 331LHN model. As we mentioned, the difference between the 331RH$\nu$ and the 331LHN is in the character of the third component of leptonic triplet, which in the later model is a new neutral left handed fermion, $N_L$, besides its right handed component, $N_R$, as a singlet under the gauge symmetry. A new symmetry was identified that transforms only the fields which are typically associated to the EWSM extended symmetry group. It was a global symmetry, $U(1)_{G}$, and the fields transforming non-trivially under it are given the following $G$-charge,
\beq
{\mathbf G}(\bar{N}_{L/R},\,\bar{u}_{3L/R}^\prime\,,d_{iL/R}^\prime,\, V^-_{\mu},\,U_\mu^0,\,\chi^0,\,\chi^-,\,\eta^{\prime 0*},\,\rho^{\prime -})=+1\,.
\label{U1G}
\eeq

Notice that, except for $N_{L/R}$, all fields that carry $G$-charge are bileptons and, again, in some sense it seems that lepton number is somehow involved in this new symmetry, without any clear identification of its role yet. The $G$-symmetry is claimed to stabilize the WIMP, which can be any electrically neutral $G$-charged field, $U^0$, $N_a$ and $\phi$, recalling that this last one is a combination of $\chi^{0*}$ and $\eta^{0}$, according to Eq.~(\ref{phi}). The non-hermitean gauge boson, $U^0$, is not interesting from de CDM point of view, since it is extremely underabundant, so that it is chosen to be heavier than the lightest neutral fermion, labeled $N_1$, and the scalar field, $\phi$. These cannot be coexisting CDM particles, then both possibilities were separately studied in Ref.~\cite{331LHNUS}. Here we face the same dilema as before in what concerns the symmetry to protect the WIMP from decay. Being a global symmetry, gravity effects can cast doubts about the WIMP stability, a point not considered in the original work. This is an important issue for the $N_1$ since the following non-renormalizable operator emerges at planck scale~\cite{singletportal} and promotes its decay into standard neutrinos,
\be
\frac{\lambda_{N_{1R}}}{\Lambda_{Pl}}\overline{N_{1R}}\,D\!\!\!\!/ f_{1L} \,\eta^\dagger + {\mbox h.c.}\,.
\label{opefNR}
\ee
Then, although the results concerning abundance and direct detection allow for a $N_{1R}$ WIMP in the model, it is not realistic since the above operator would demand it to be too much lighter (less than few keV~\cite{singletportal}) than the TeV scale pointed in Ref.~\cite{331LHNUS} in order to stay stable enough.

In this framework, there seems to remain only one interesting WIMP in this model, the scalar $\phi$. We have to look for the main dangerous effective operator that breaks $U(1)_G$ at Planck scale and still preserves the gauge symmetry. Such operators were mentioned in that work having a grand unification scale in mind, they would give Majorana mass for the neutrinos and were discarded once they explicitly violate $U(1)_G$,
\be
\frac{y^\prime_{ab}}{\Lambda_{GUT}}\overline{f_{aL}^c}\, \chi^*\,\chi^\dagger f_{bL} +\frac{y^{\prime\prime}_{ab}}{\Lambda_{GUT}}(\bar{f_{aL}^C}\chi^*)(\eta^\dagger f_{bL})+\mbox{h.c.}\,.
\label{efopfi}
\ee
However, they must be considered at Planck scale. Differently from the 331RH$\nu$ model, where the $Z_2$ symmetry is supposed to be remnant  from a gauge theory (not broken by gravity effects), in the 331LHN such an argument does not work for the global $U(1)_G$ and the second operator in Eq.~(\ref{efopfi}) would lead to fast decay of $\phi$ into neutrinos, unless the dimensionless coupling, $y^{\prime\prime}$ be unnaturally tiny, which would definitely turn this model awkward in what concerns the CDM problem.

That is not the final answer though. A smart solution exists for the choice of the correct symmetry that guarantees the existence of a WIMP in this model, which is based on the observation that the new neutral fermions and the bileptons carry the ``wrong'' lepton number~\cite{tuly}. Namely, the third component of the leptonic triplet carries a lepton number which is generally the opposite of its partners in the multiplet in the 331RH$\nu$ model, and it can be taken as null in the 331LHN~\cite{dongsoa}, while the new quarks, some gauge bosons and scalar fields also carry lepton number in contrast to EWSM similar fields. This possibility lead to an enlargement of the 331 gauge group by
an extra $U(1)$ gauge symmetry, called 3311 model which, after spontaneous breakdown, still has a discrete $Z_2$ symmetry that remains unbroken and coincides with the R-parity symmetry of supersymmetry~\cite{3311}, 
\be
W=(-1)^{3(B-L)+2s}\,,
\label{W-parity}
\ee
called W-parity (W referring to wrong lepton number), where $B$ is the baryon number, $L$ is the lepton number and $s$ is the spin of the field.
Then, by assuming this unbroken W-parity and considering the quantum numbers assigned to the 331LHN fields we get the odd transforming fields under this $Z_2$~\cite{3311,FJZlinha,wimpyDR},
\be
W(N_{L/R},\,d_i^\prime ,\,u_3^\prime ,\,\rho^{\prime \pm} ,\, \eta^{\prime 0},\, \chi^0 ,\, \chi^\pm, \, V^\pm,\, U^0)=-1\,.
\label{Wnumbers}
\ee
Being an exact symmetry at low energies, it guarantees the stability of the lightest particle that transforms nontrivially under it, which means that either $\phi$ or $N$ can be a (non-concomitant) realistic WIMP.
Those two particles were extensively studied in this way, and besides the simple analysis concerning their relic abundance and direct detection~\cite{331LHNUS}, several complementary interesting analyses were pursued that we summarize below.

\begin{figure}[!h]
\centering
\includegraphics[width=0.7\columnwidth,angle=-90]{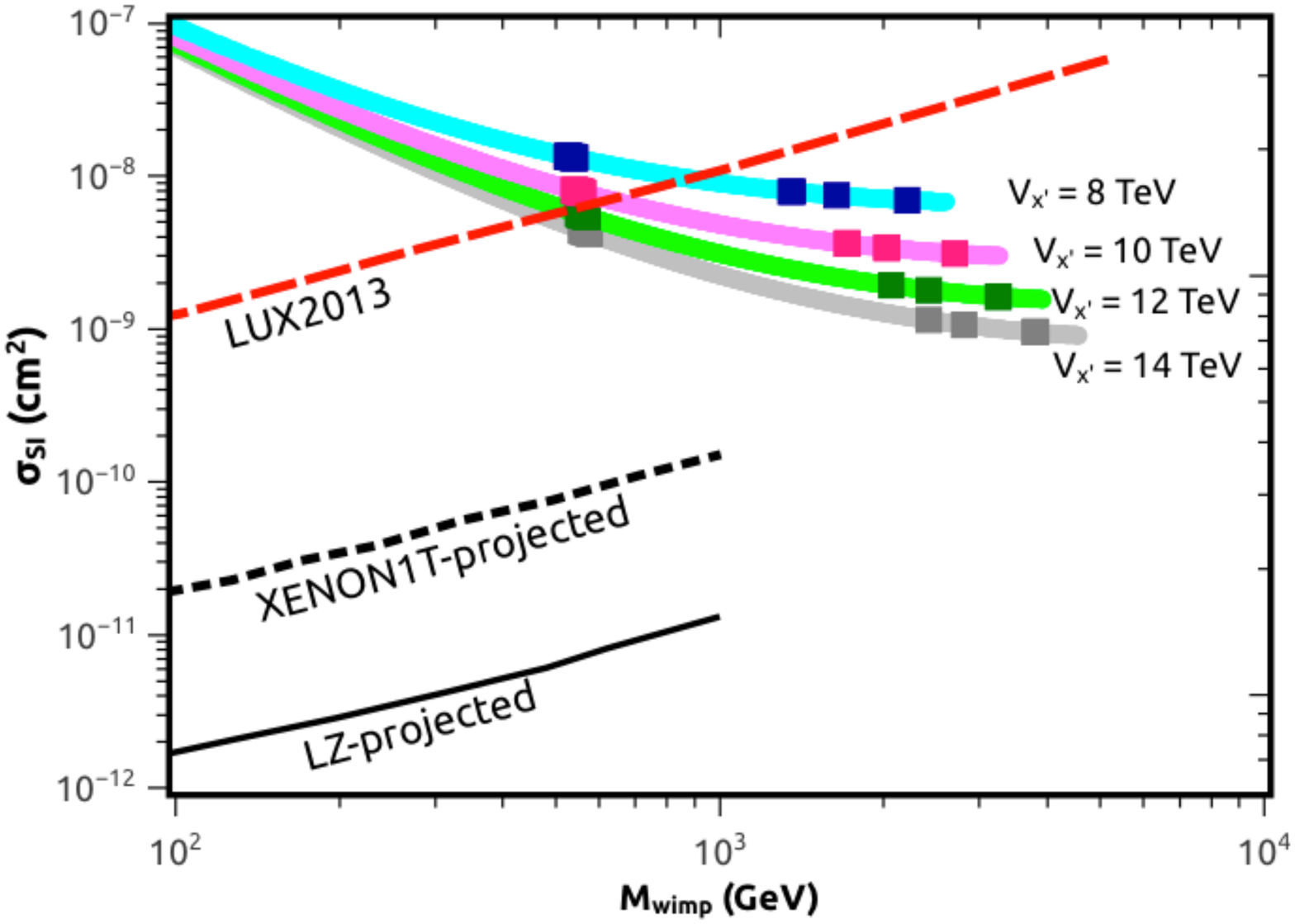}
\caption{WIMP-nucleon spin independent scattering cross section for $v_{\chi}=8,10,12,14$~TeV. The dark points delimit the parameter space that yields the right abundance in accordance with Planck~\cite{planck}, Eq.~(\ref{limitsDM}). The dashed red (black) curve is the LUX2013~\cite{LUX} (XENON1T-projected~\cite{xenon1t}) limit, while the solid black line is for the projected LZ limit~\cite{LZ}. Figure extracted from Ref.~\cite{FJMZL} .}
\label{fig:3}
\end{figure}

When the complex scalar field, $\phi$, is chosen to be the lightest W-particle in the 331LHN model, it was shown to give the right abundance, scape direct detection and could also explain the gamma ray emission from the galactic center~\cite{DMH}, as inferred by data from Fermi-LAT satellite~\cite{gcemission}. This can be achieved for a WIMP mass between $25\lesssim m_\phi \lesssim 40$~GeV, when this WIMP predominantly annihilates into $b\bar b$ (more than $50\%$), as reproduced in Fig.~\ref{fig:2} below. 
\begin{figure}[!htb]
\centering
\includegraphics[width=0.7\columnwidth, angle=-90]{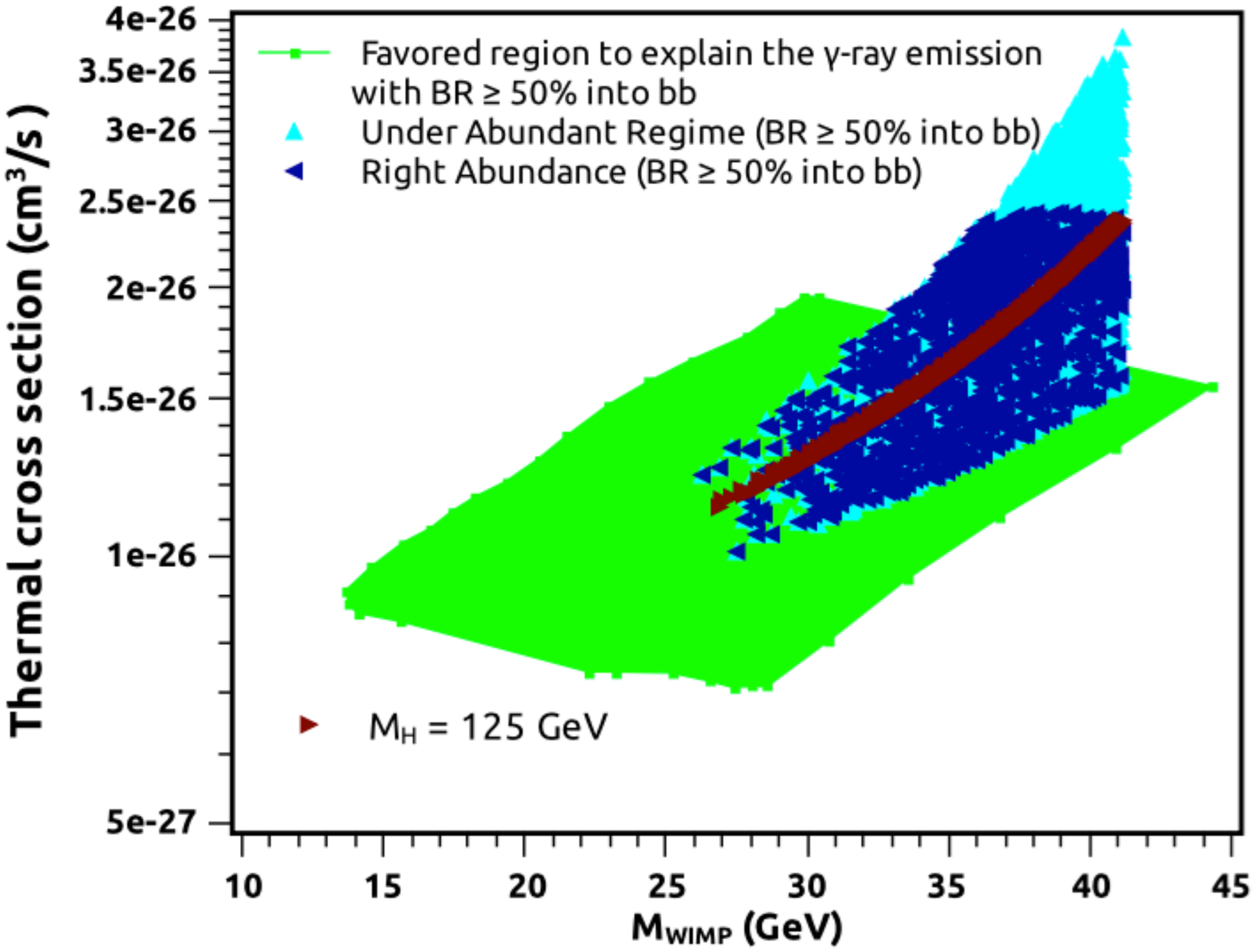}
\caption{Annihilation cross section as function of $M_{\phi}$ for $BR(b\bar{b}) \geq 50\%$. The green region represents the favored region by the gamma-ray emission detected by Fermi-LAT in the galactic center~\cite{gcemission}. Dark (light) blue  points refer to the case where the WIMP provides the correct abundance (under abundant). All (dark+light) blue points are for $110\ \mbox{GeV} \leq M_H \leq 150\ \mbox{GeV}$, while the brown points are for $M_H = 125$~GeV. Correct abundance means $ 0.098\leq \Omega h^2 \leq 0.122$ while the under abundant regime is for ($0.01 \leq \Omega h^2 \leq 0.098$). Figure taken from Ref.~\cite{DMH}.}
\label{fig:2}
\end{figure}
Nevertheless, in the same work, this result was confronted with Higgs physics from LHC. Although such a light WIMP could solve the galactic center gamma ray emission in this model, it does not comply with Higgs decay as seen by LHC~\cite{ATLASCMS}, since the model predicts a too high branching ratio into light WIMPs (whose mass is less than 60~GeV), close to 90$\%$, while predicting a too low branching ratio into two photons~\cite{DMH}. Also, the scalar WIMP mass was further constrained in this model by considering bounds on the $Z^\prime$ mass, pushing the symmetry breaking scale of 331LHN to $v_{\chi^\prime} \gtrsim 10$~TeV, which contrasted with recent LUX results on direct detection~\cite{LUX}, implies a lower bound to $\phi$ mass, $m_\phi \gtrsim 500$~GeV~\cite{FJMZL}, as can be seen in Fig.~\ref{fig:3}.
Notice from that plot that XENON1T projected limit~\cite{xenon1t} will be able to exclude the scalar dark matter mass range below $1$~TeV in the 331LHN model.

\begin{figure*}[!t]
\centering
\includegraphics[width=0.7\columnwidth]{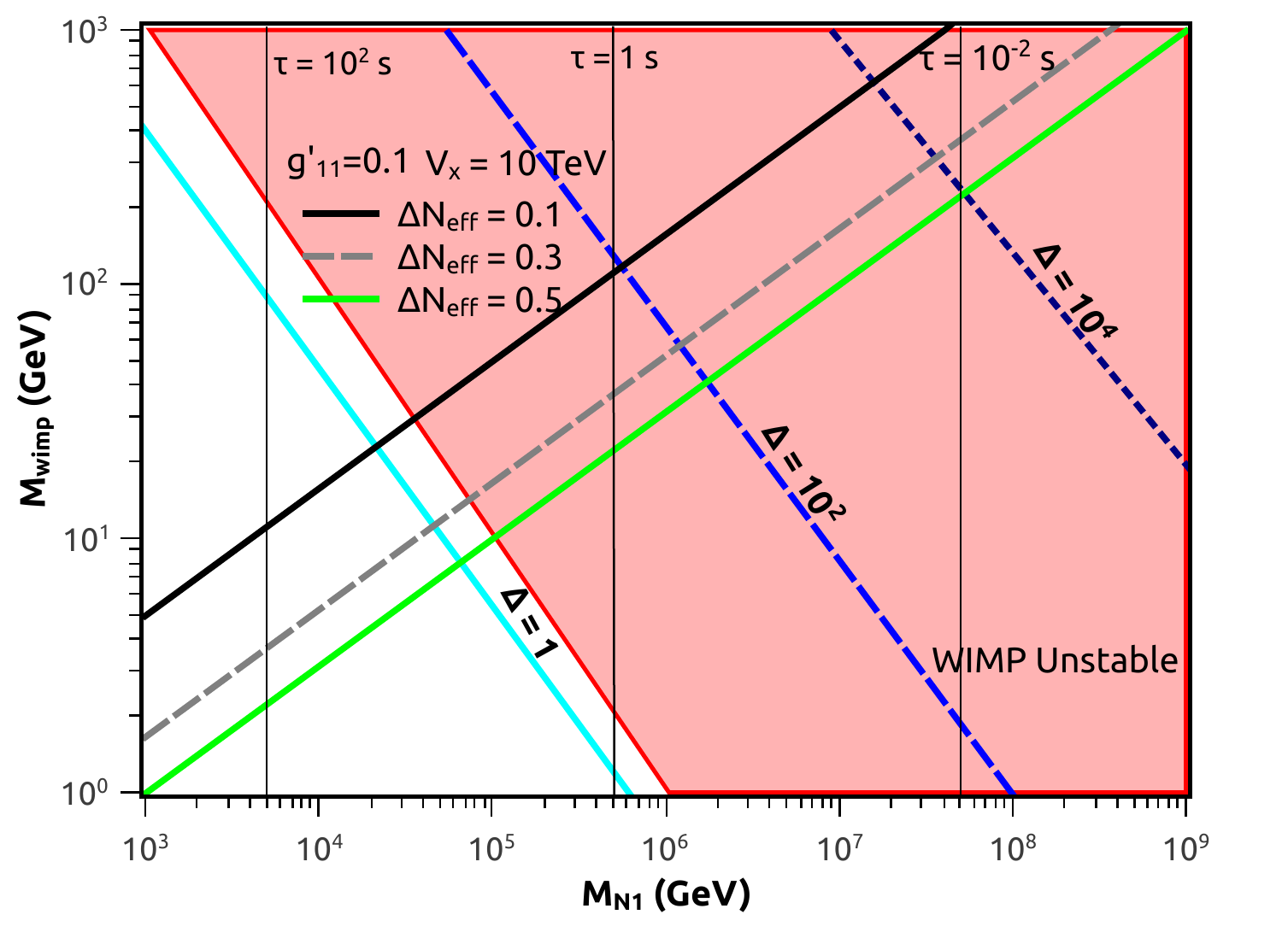}
\caption{The ``mother-daughter'' particle mass parameter space, for $g_{11}^\prime=1$. The vertical lines indicate constant values of the mother particle lifetime. The diagonal lines indicate the induced variation in the number of effective relativistic degrees of freedom $\Delta N_{\rm eff}$ and the entropy dilution factor $\Delta$ needed to suppress the mother particle relic density. The cyan $\Delta=1$ line corresponds to standard cosmology without any entropy dilution needed. Here, $v_{\chi}^{\prime}=10$~TeV and the red shaded region induces the WIMP decay. Figure extracted from Ref.~\cite{wimpyDR}.}
\label{fig:4}
\end{figure*}

Another interesting feature of this 331LHN model lies on the possibility to address the problem of extra radiation in the CMB spectrum, the so called Dark Radiation~(see Ref.~\cite{DR} and references therein). A number of experiments have shown a mild but convincing preference for an effective neutrino number higher than the standard $N_{eff}=3.046$ (a nice compilation of results is presented in Ref.~\cite{DRexp}). The central value excess is around $\Delta N_{eff}\approx 0.5$ when considering the combined data from Planck + SPT + WMAP + ACT +  $H_0$~\cite{planck}, and can be explained in the context of a heavy particle decaying into relativistic WIMPs at radiation dominated era~\cite{DRdecay}. This explanation fits well in the context of 331LHN model, since the lightest heavy neutral fermion, $N_1$, can play the role of the mother particle, decaying into $\phi$ plus an active neutrino with parameters appropriate to produce only a fraction of relativistic WIMPs, remarking that this fraction must not exceed $0.01\%$ of the whole number of WIMPs so as to not jeopardize structure formation~\cite{DRFH}. This approach was followed in Ref.~\cite{wimpyDR} and has been shown to be feasible for the 331LHN model while keeping consistency with the above mentioned constraints on $v_{\chi^\prime} \gtrsim 10$~TeV and $m_\phi  \gtrsim 500$~GeV~\cite{FJMZL}. These results are evident from Fig.~\ref{fig:4}, where the mother particle is the lightest neutral fermion and the daughter particle is the scalar WIMP, which couple to each other with coupling strength $g_{11}^\prime$ originating from the Yukawa coupling in Eq.~(\ref{yukawa2}). In this plot $f$ is the fraction of WIMP that is produced relativistically from $N_1$ decay and $\Delta$ is a suppression factor that can account for a possible high relic number density of the mother particle, which may be the case if $g_{11}^\prime$ is to small, possibly pushing the model into the region where $\phi$ ceases to be stable (the red shaded area in the plot).

\begin{figure}[!htb]
\centering
\includegraphics[width=0.7\columnwidth,angle=-90]{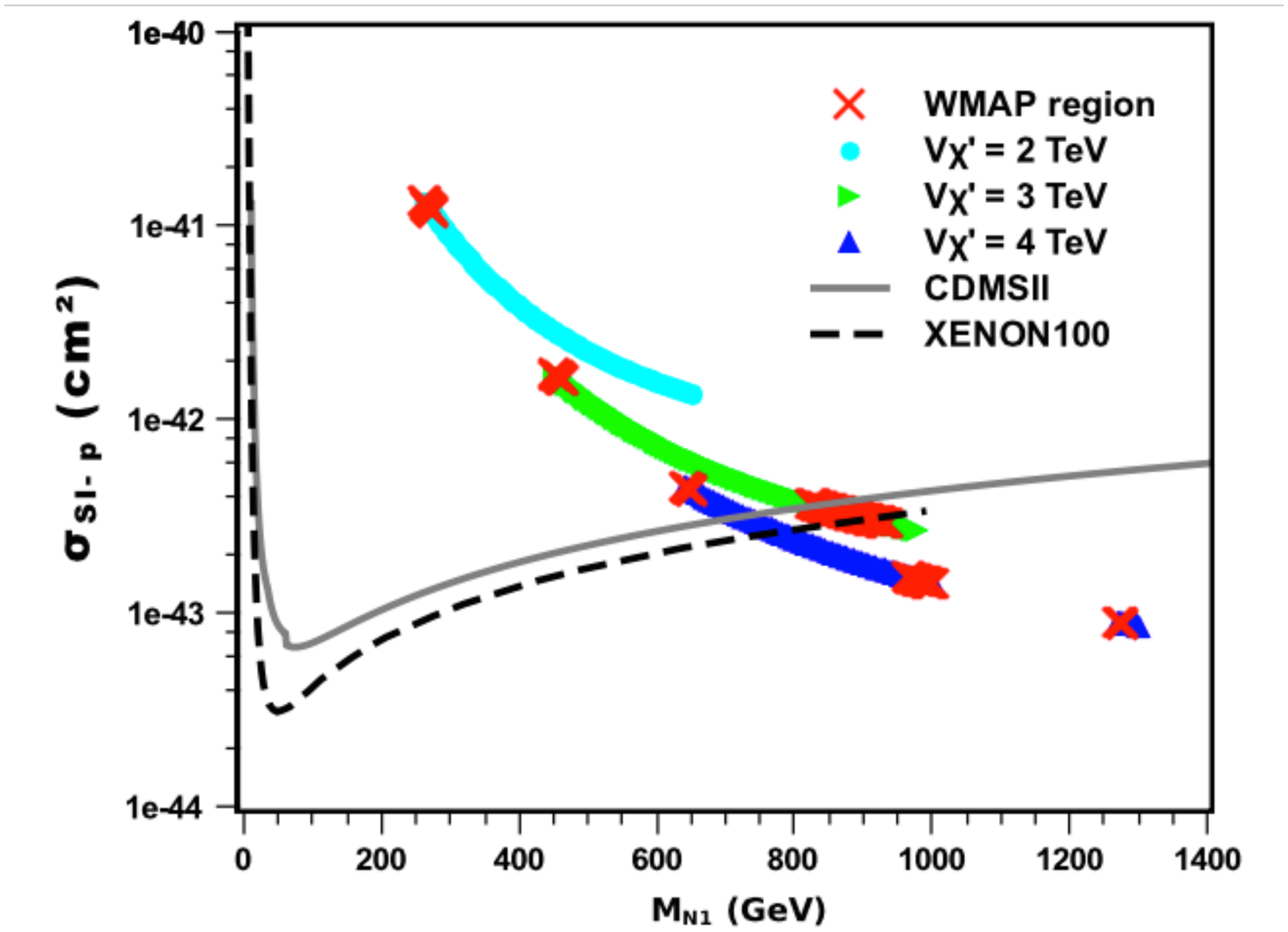}
\caption{The WIMP-proton cross section for $N_{1}$. From top to bottom, the curves represent the variation of $v_{\chi^\prime}$ in the range $2$~TeV$\leq v_{\chi^\prime} \leq 4\ \mbox{TeV}$. The data used in the exclusion curves were obtained using \cite{XENONtools}. Figure extracted from Ref.~\cite{331LHNUS}.}
\label{fig:5}
\end{figure}

\begin{figure}[!t]
\centering
\includegraphics[scale=0.7]{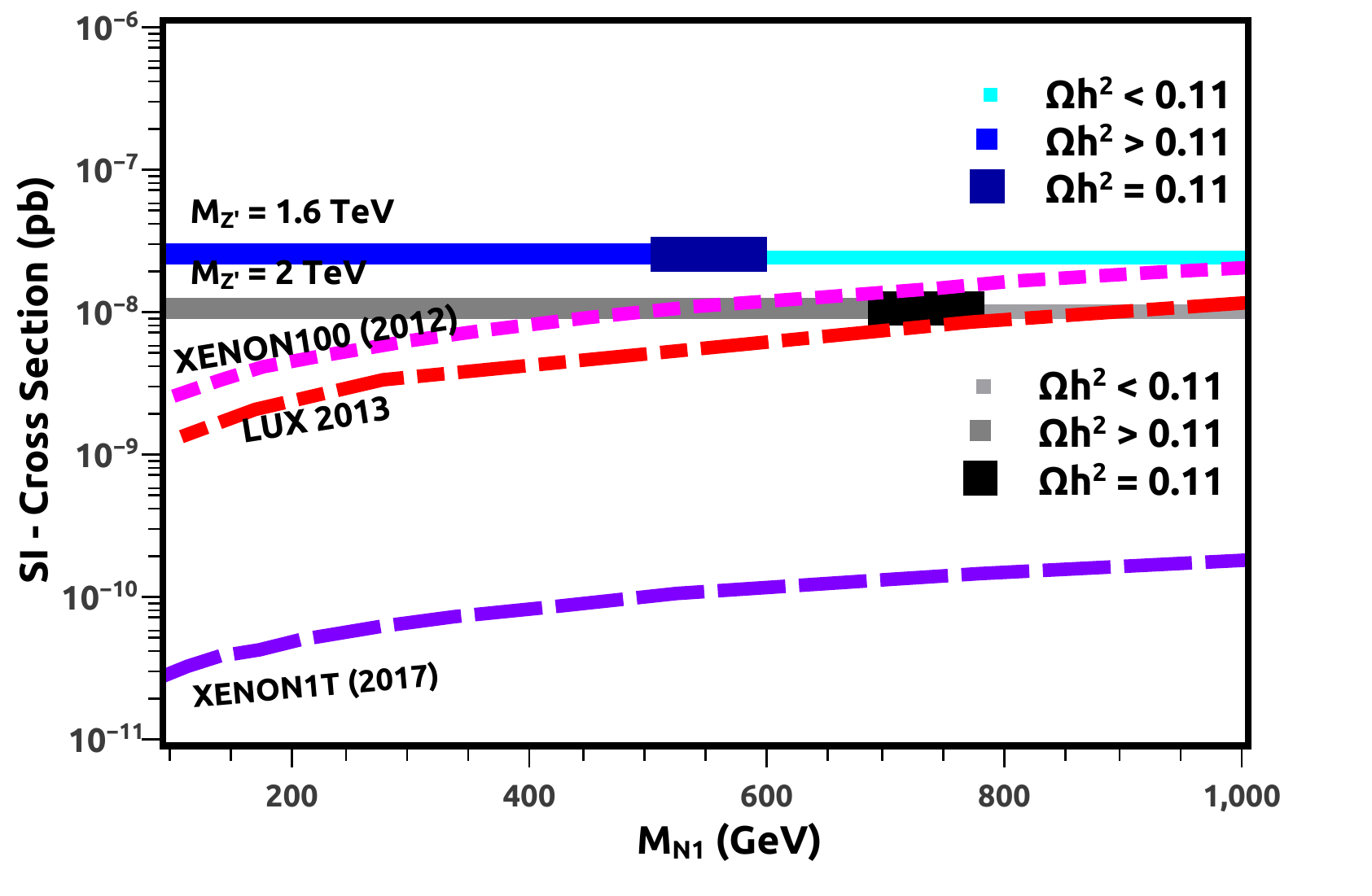}
\caption{$N_1$ spin independent scattering cross section off nuclei. Figure extracted from Ref.~\cite{FJZlinha}.}
\label{fig:6}
\end{figure}

The other interesting possibility is when one of the new neutral fermions, $N_a$, is the lightest odd W-parity, becoming the CDM candidate. In this case, although there are regions in the parameter space compatible with the observed relic abundance for $m_N\approx 200$~GeV, the WIMP is pushed into masses around 1~TeV when confronted with direct detection experiments, implying a 331LHN symmetry breaking scale, $v_{\chi^\prime}\gtrsim 4$~TeV~\cite{331LHNUS}, as can be seen from Fig.~\ref{fig:5}. 
An interesting study considering the role of $Z^\prime$ in the WIMP-nucleon scattering amplitude, translated into bounds in the cross section coming from direct detection experiments, mainly the recent LUX results~\cite{LUX} which, together with constraints from CDM relic abundance, forced $M_{Z^\prime}\gtrsim 2$~TeV for a WIMP mass around 1~TeV~\cite{FJZlinha}, as shown in Fig.~\ref{fig:6}. 
This further pushes the lower bound on the breaking scale, $v_{\chi^\prime}\gtrsim 5$~TeV.
Observe that these constraints are an important outcome concerning this 331LHN model, since the bounds on $M_{Z^\prime}$ obtained in Ref.~\cite{yara} do not apply in the case $m_N\leq M_{Z^\prime}/2$, where the main $Z^\prime$ decay channel into charged leptons is suppressed due to this preferred invisible channel into two WIMPs. It is another complementary connection from dark matter that should be taken into account in the specific case where the heavy neutral fermion is the CDM candidate in 331LHN~\cite{FJZlinha}. 
Similar results, for both CDM candidates, were obtained in a  phenomenological study of the 331LHN model with an extra gauge $U(1)$ group broken to the W-parity symmetry~\cite{3311phen}~\footnote{The WIMP candidates are exactly the same since this model is a simple enlargement of the gauge group by an abelian factor.}.

\section{Conclusions}
\label{sec4}

We have revisited the electroweak gauge extensions of the SM, the 331 models with neutral fermions, considering their ability to provide a good CDM candidate, more specifically, a WIMP. Two such models were considered, the 331RH$\nu$~\cite{jcap} and the 331LHN~\cite{331LHNUS}, differing by the fact that in the first model the neutral fermion is the right handed partner of the left handed neutrino, while in the second one it is a new kind of neutral fermion that may not carry lepton charge. Both models possess a complex scalar CDM candidate, while the 331LHN offers an additional non-concomitant candidate, the new neutral fermion.
The properties that guarantee the existence of a WIMP were analyzed, and essentially the main problem in these models has been finding the appropriate symmetry to protect the WIMP from decaying. 

Such stability, as discussed, is not guaranteed for the 331RN$\nu$, since the proposed symmetry is spontaneously broken. While lepton number seems to play some role in keeping the WIMP, $\phi$ in this model, stable at tree level, higher dimensional effective operators imply the WIMP decay due to gravitational effects, with a lifetime $\tau\approx 10^{43}$~s, big enough to consider it stable. Constraints from $Z^\prime$ mass push the symmetry breaking scale to $v_{\chi^\prime}\gtrsim 5.5$~GeV and, according to the results shown in Ref.~\cite{jcap}, a scalar WIMP of mass around 1~TeV would be compatible with relic CDM abundance and direct detection. 

Considering the 331LHN model, the global $U(1)_G$ symmetry proposed in Ref.~\cite{331LHNUS} was not suitable to provide the WIMP stability, once the same effective operators that respect the symmetries of the model and are employed to generate Majorana mass for the neutral fermions, would be induced by gravitational effects in what concerns lepton number violation. Differently from 331RH$\nu$ model, such operators lead to fast WIMP decay, unless some huge fine tuning is claimed. However, a new discrete symmetry called W-parity, similar to R-parity in supersymmetric models, was implemented in this model when the neutral fermion in the leptonic triplet does not carry lepton number~\cite{3311}. It was first used as a remnant of a specific gauge symmetry in Ref.~\cite{3311,3311phen}, and employed afterwards without worrying which gauge symmetry may be behind its survival at low energies~\cite{FJZlinha,wimpyDR,FJMZL}. In this case, all results obtained before remain valid for the 331LHN model~\cite{331LHNUS,DMH}, since no relevant changes have affected the WIMP identity and properties. 

The interesting outcomes of 331LHN CDM model reveal that the scalar WIMP, $\phi$, which could have a large mass range~\cite{331LHNUS}, could offer an explanation for the excess of gamma ray emission from the galactic center when $25\lesssim m_\phi\lesssim 40$~GeV. That was shown to be unrealistic when confronted with complementary phenomenology from recently observed Higgs physics~\cite{DMH}, since the Higgs decay was predicted to be more than $90\%$ into two WIMPs if $m_\phi\lesssim 60$~GeV, completely ruled out by LHC data~\cite{ATLASCMS}.
Further analysis has shown that $m_\phi\gtrsim 500$~GeV, and $v_{\chi^\prime}\gtrsim 10$~TeV when the bounds over the mass of the new neutral gauge boson, $Z^\prime$, were considered~\cite{FJMZL} in conjunction with recent constraints of CDM direct detection experiment LUX~\cite{LUX}.  Besides, the model accommodates the right content to address a solution to the dark radiation problem through the out-of-equilibrium production of a fraction of the WIMP as relativistic species~\cite{wimpyDR}.

Finally, when the WIMP from 331LHN model was chosen to be the lightest heavy neutral fermion, $N_1$, the first analysis made already evident that it should have a mass around 1~TeV to be consistent with relic abundance and direct detection experiments~\cite{331LHNUS}. The bounds on the symmetry breaking scale are weaker in this case, though. Since existent limits from LHC physics on the $Z^\prime$ mass~\cite{yara} may not apply to this model for $M_N\lesssim 1$~TeV, in which case $Z^\prime$ would prefer to decay invisibly into two WIMPs, a bound over the $Z^\prime$ mass was found in this specific case by considering LUX results, obtaining $M_{Z^\prime}\gtrsim 2$~TeV, and thus, $v_{\chi^\prime}\gtrsim 5$~TeV~\cite{FJZlinha}. It is worth to be mentioned that this result truly complements previous results on $Z^\prime$ mass~\cite{yara} and shows the importance of CDM approach to electroweak precision phenomenology.

\noindent {\bf Acknowledgments:}\\
The author feels deeply grateful to his collaborators, Alex Dias, Carlos Pires and Farinaldo Queiroz. This work was supported by the Conselho Nacional de Desenvolvimento Cient\'{\i}fico e Tecnol\'ogico (CNPq).

%
\noindent{\bf References} \\

\end{document}